\documentstyle[prl,aps]{revtex}
\begin{document}

\draft

%{CWRU-P15-97}
%{Imperial/TP/97-98/3}
%{CERN-TH/97-273}

\title{
Sweeping Away the Monopole Problem
}


\author{
G. Dvali$^{\dag \P}$, 
Hong Liu$^{\ddag}$, and, Tanmay Vachaspati$^{\S}$
}
\address{
$^{\dag}$TH-Division, CERN, CH-1211 Geneva 23, Switzerland.\\ 
$^{\P}$ICTP, POB 586, 34100, Trieste, Italy.\\
$^{\ddag}$Imperial College, Theoretical Physics, Prince Consort Road,
SW7 2BZ London, England.\\
$^{\S}$Physics Department,
Case Western Reserve University,
Cleveland OH 44106-7079, USA.
}

%\date{\today}

\twocolumn[

\maketitle

%\tightenlines

\begin{abstract}
\widetext

We propose a new solution to the cosmological monopole problem
in which domain walls sweep away the magnetic monopoles and
subsequently decay. The solution does not require extensive
fine tuning or model building - it works for the prototype 
$SU(5)$ Grand Unification model. More generally, it shows that
defect interactions can lead to ``defect erasure'' in phase
transitions and that this can be relevant to early universe
physics.

\end{abstract}

%insert suggested PACS numbers in braces on next line
\pacs{}

]

\narrowtext

All known attempts to unify the fundamental forces of
nature predict the existence of magnetic monopoles \cite{avps}.
In fact, in a cosmological context, all such attempts predict
the existence of too many magnetic monopoles \cite{preskill}. 
This is the monopole problem.

The monopole problem has at least three known solutions. The first
is the inflationary solution to the monopole problem \cite{guth}
whereby the universe inflates and dilutes the monopole density
to acceptable levels.
The second is called the Langacker-Pi mechanism \cite{lpi}
in which monopoles and antimonopoles get connected by strings 
which draw them together, leading to annihilation. The third
mechanism \cite{gs} relies on non-restoration of the grand
unified symmetry and so there never was a phase transition in
which monopoles were produced.

Here we show that there is yet another mechanism for solving
the magnetic monopole problem that is economical -
does not require complicated particle physics model building -
and does not suffer from fine tuning of the kind encountered
in generic inflationary models. Indeed, the simplest concrete 
realization of the model is none other than the prototype
Grand Unified $SU(5)$ model that first inspired the inflationary
solution.

The basic idea of this new mechanism is quite simple. The
phase transition that produces magnetic monopoles, also
produces domain walls. The domain walls move through space
and sweep up all the monopoles. When a monopole encounters
a wall, it unwinds and dissipates and, in this way, the walls
sweep away the monopoles from the universe.
The alert reader would have realized at once that a danger 
in this scheme is that the monopole problem may have been replaced
by a domain wall problem. This would be true if the walls
were stable. However, as we discuss below, the walls
in this scheme can be unstable at a lower energy scale and hence
collapse and go away. The reason for the instability is that 
the discrete symmetry responsible for the walls is chosen to
be approximate, or, in another rendering of the scenario, 
instanton effects violate the $Z_2$ symmetry and destabilize
the walls. 

The constraints on the model come from requiring that domain walls 
do not dominate the universe but live
long enough to solve the monopole problem. As we shall see, these 
constraints can be met without any severe fine tuning.

Let us now describe a concrete realization of the scenario.
We consider the $SU(5)$ Grand Unified model with an adjoint, $\Phi$,
(and a fundamental) scalar field. The Higgs potential for $\Phi$ 
is the standard \cite{chengli}
$$
V(\Phi ) = - {1 \over 2}m^2 {\rm Tr}\Phi^2 + 
                {h \over 4}({\rm Tr}\Phi^2)^2 +
                {\lambda \over 4}{\rm Tr }\Phi^4 + 
                {\gamma \over 3} m {\rm Tr} \Phi^3
$$
where, $\gamma$ is the dimensionless parameter that characterizes
the explicit violation of the $Z_2$ symmetry: $\Phi \rightarrow
-\Phi$. 

Consider the case $\gamma =0$, in which case the spontaneous
symmetry breaking 
\begin{equation}
SU(5)\times Z_2 \rightarrow 
[ SU(3)_c \times SU(2)_L \times U(1)_Y ]/Z_6 
\label{symbreak}
\end{equation}
occurs when $\Phi$ acquires a vacuum expectation value (vev)
$
\Phi_0 = v {\rm diag} (2,2,2,-3,-3)/\sqrt{30}  
$
where $v = m /\sqrt{\lambda '}$ with $\lambda ' \equiv h +7\lambda /30$.
To pick out this direction for the vev, we need the following constraints 
on the parameters in the Higgs potential:
$ \lambda > 0$, $h > - 7 \lambda /30$  ({\it i.e.} $\lambda ' > 0$).
The vev $\Phi = -\Phi_0$ also leads to the symmetry breaking 
in (\ref{symbreak}). The two discrete vacua $\Phi = \pm \Phi_0$ are 
degenerate due to the exact $Z_2$ symmetry.

If $\gamma$ is non-zero but small enough to lead to the symmetry
breaking in (\ref{symbreak}), the potential will continue
to have two discrete minima which will now be non-degenerate. 
(Our scheme should also work for domain walls interpolating between
vacuua with different symmetries but, for simplicity, we do not
consider this possibility.)
%(For simplicity, we have chosen not to
%entertain the possibility of other discrete local minima with different
%symmetries {\it eg.} $SU(4)\times U(1)$.) 
%Our scheme should work for
%domain walls interpolating between vacuua with different symmetries
%as well.)
These minima will survive 
as long as the cubic term in the potential is small compared to the other 
terms and so leads to an order of magnitude constraint on $\gamma$: 
it should be less than the other coupling constants in the
model.  Denote the vev of $\Phi$ in the lower energy minimum by 
$\Phi_+$ and that in the higher energy minimum by $\Phi_-$. In the 
limit of vanishing $\gamma$, $\Phi_\pm \rightarrow \pm \Phi_0$.

It is well known that the symmetry breaking (\ref{symbreak})
leads to magnetic monopoles and $Z_2$ domain walls. What is less
appreciated is that, if $\gamma \ne 0$, the symmetry breaking
still leads to the formation of cosmological domain walls
that interpolate between the two non-degenerate discrete minima
denoted by $+$ and $-$.
In a cosmological setting, if the $Z_2$ symmetry is exact ($\gamma =0$), 
we know that regions of the $+$ and $-$ vacuua will percolate 
\cite{stauffer} and the domain wall network will consist of 
an infinite domain wall and (very few) small isolated walls
\cite{harvey,tvav}. When 
$\gamma \ne 0$, the universe relaxes into the $+$ vacuum 
with higher probability than in the $-$ vacuum leading to
``biased'' domain wall formation \cite{tvformn}. 
If we denote the probability of a spatial domain being in the 
$+$ vacuum by $P_+$ and that of being in the $-$ vacuum by $P_-$, we have
$P_- = P_+ {\rm exp} [ - {{\Delta F_V} / T_c} ]$,
where, $\Delta F_V$ is the free energy difference between two domains
of volume $V$ of the $\pm$ vacuua, and, $T_c$ is the freeze-out temperature 
of the domains during the phase transition \cite{zurek}. The free energy
difference is given by 
$\Delta F_V \simeq 2\gamma m v^3 V /3\sqrt{30}$.
If $P_-$ is not too small, the walls will still percolate. For example, 
if percolation on a cubic lattice is a good description of the phase 
transition, the walls will percolate if $P_\pm > 0.31$. Using $P_- = 1-P_+$, 
and requiring wall percolation imposes a constraint, 
$\Delta F_V < 0.8 T_c$, which translates into the constraint:
$\gamma \lesssim 7 T_c /m V v^3$.
For weakly first order and second order phase transitions,
on dimensional grounds, $V \sim m^{-3}$, $T_c \sim m$ and for values of
$h$ and $\lambda$ that are not too extreme, we have $v \sim m$. 
Hence the constraint on $\gamma$ is quite mild and no fine tuning 
is needed to get infinite walls to be produced. For strongly first order
transitions, $V$ can be as large as the horizon volume at the Grand
Unified phase transition and the constraint on $\gamma$ is very
strong. Here we will only consider the weakly first order or second
order phase transition for which $Vm^3$ is not too large.

The energy difference across the walls for $\gamma \ne 0$
implies a force that drives the walls into the $-$ regions. For 
example, a straight infinite wall would be pressure driven 
so that the volume in the higher energy $-$ vacuum gets smaller. 
However, in a cosmological setting and at early times, the force of tension, 
$\sigma /R$ (where $\sigma = 4m^3/3\lambda '$ 
%note - corrected
is the energy per unit area 
of the wall and $R$ is the radius of curvature) can be large
compared to the pressure difference, $\Delta F_V /V$, across the wall.
The radius of curvature of the walls $R$ is of order $T_c^{-1}$
for second order phase transitions. Hence, the pressure contribution 
to the dynamics of the walls is sub-dominant provided: 
\begin{equation}
\gamma \lesssim {{10 \sqrt{\lambda '}} \over {R m}} \ ,
\label{tension}
\end{equation}
% note - corrected above equation (sqrt included)
which gives $\gamma \lesssim 10 \sqrt{\lambda '}$.
With this constraint satisfied,
the wall evolution is just as in the exact $Z_2$
symmetry case where there is no pressure difference. As the walls 
are two dimensional surfaces moving in three spatial dimensions and 
are infinite in extent with no spatial symmetries, they will 
sweep out the entire volume of the universe in a time 
$ \tau \sim R / v_w$
since $R$ is also the inter-wall distance and where $v_w \sim c$ is 
the wall velocity. 

As the walls sweep the universe, they also sweep up the monopoles.
Now what happens when a monopole hits a domain wall? Here there
are two parameter dependent possibilities that we must discuss
separately. The first is when $\Phi =0$  and the other is when 
$\Phi \ne 0$ inside the wall. To see that both cases are possible,
consider the Higgs potential when we restrict $\Phi$ to lie
along the diagonal,
$a \lambda_3 + b \lambda_8 + c \tau_3 + v Y$,
where $\lambda_3$ and $\lambda_8$ are matrices from the $SU(3)_c$ Cartan
subalgebra, $\tau_3$ is the weak isospin and $Y$ is the hypercharge 
generator, all matrices being normalized to unity.
Usually one assumes that since $\Phi \rightarrow -\Phi$ across the
wall, we must have $\Phi =0$ at the center of the wall and the $SU(5)$ must 
be restored on the wall. However for a wide range of parameter space this
is not the case because some other component(s) of $\Phi$ 
(which vanish in the vacuum) can pick up a vev and break the 
gauge symmetry inside the wall in a different fashion.  This can be 
simply understood by examining the linearized Schrodinger
equation for small excitations $\epsilon = \epsilon_0 (x) e^{-i\omega t}$
($\epsilon$ is either the $a,b$ or $c$ component) in the wall background
$$
\left [-\partial_x^2 + (-m^2 + ({\bar v}(x))^2 (h + \lambda r)) \right]
\epsilon_0 = \omega^2 \epsilon_0
$$
where, the wall is taken to lie in the $x=0$ plane and
for $\epsilon$ being in the $a,b,c$ directions we have 
$r = 2/5, ~ 2/5 ~ {\rm and} ~ 9/10$ respectively. 
The function ${\bar v}(x)$ is the profile of the wall which 
for a planar infinite wall can be approximated by a kink solution
${\bar v}(x) = v {\rm tanh} (mx/\sqrt{2})$.
When $\lambda = 12 \lambda '$ and $r=2/5$, the
Schrodinger equation is identical to the equation obtained when
solving for perturbations around the kink solution \cite{rajaraman}.
Then there is a zero mode corresponding to the translation of the
kink. From this we deduce that the domain wall will have $\Phi =0$
at the center only if $\lambda ' < \lambda /12$ 
as there are no bound state (negative $\omega^2$) solutions to the 
Schrodinger equation if this condition holds. 

We must comment here that even if one forgets the monopole problem, 
from the cosmological viewpoint one prefers the range of parameters 
for which $\Phi$ vanishes on the wall. The reason is that if
$\Phi$ is nowhere zero in space, then the domain walls may never
collapse even though the $Z_2$ is only approximate\cite{2piwall}.
To see this, let $b$ be a component that is non-zero 
on the wall and define a complex number $\psi = {\bar v} + ib$
with the boundary conditions $\psi = \pm v$ on opposite sides of
the wall.
This configuration is equivalent to the winding of the phase of $\psi$
by $\Delta =  \pi$ through the wall. Since the expectation value
$|\psi|$ is nowhere zero, the winding of the phase is well-defined and
one can distinguish walls with $\Delta = \pm \pi$ when one crosses
them in one direction. Now, walls with opposite winding can 
annihilate each other and disappear, but pairs with the same winding 
cannot. So when two 
neighboring walls with equal winding are pushed towards each other 
due to the pressure differences, they cannot annihilate, but instead 
form a ``bound-state'': a $\Delta = 2\pi$ domain wall\cite{2piwall}. 
Although not truly stable (these can decay via quantum nucleation
of holes\cite{axion}) they can be stable for all practical purposes if 
the expectation value of the Higgs field in the core is large. 
This consideration disfavors (but does not exclude) the range of 
parameters for which $\Phi \ne 0$ inside the wall.

If $\Phi =0$ inside the wall, the full $SU(5)$ symmetry is restored 
there. If, however, $\Phi \ne 0$ inside the wall, the symmetry inside
the wall is not the full $SU(5)$ but only a subgroup which is different
from the unbroken subgroup in the vacuum. We expect that our mechanism
will work as long as the symmetry inside the wall is large enough so
that monopoles can unwind there. To keep the discussion simple, 
we will only consider the $\Phi =0$ case here.

The interaction of monopoles and domain walls has not
been investigated in detail but there are strong indications
that the monopoles will unwind on entering the wall where the
full $SU(5)$ symmetry is restored. These indications are as 
follows:
%\begin{enumerate}
%\smallskip

(i) There is an attractive force between the monopoles
and the walls since monopoles can save the expense of having to
go off the vacuum in their core by moving on to the wall.
So the monopoles can form bound states with
the walls\footnote{We observe here that the domain wall and
monopole bound state can lead to a classical realization of
a D-brane if the $SU(3)_c$ symmetry group further breaks to
$Z_3$ since now the monopoles bound to the walls will be
connected by strings. Related constructions may also be found in
\cite{trodden}.}. 
Then, as there is no topological obstruction to the
unwinding of monopoles on the wall, the monopoles on the wall
can continuously relax into the vacuum state.

%\smallskip

(ii) The investigation of a similar system - Skyrmions
and walls - has been dealt with in full detail in Ref. \cite{piette}.
These authors find that the Skyrmion hits the wall, sets up
traveling waves on the wall and dissipates.
They also find that, even though it is topologically possible
for the Skyrmion to penetrate and pass through the
domain wall, this never happens. They attribute their finding to the 
coherence required for producing a Skyrmion. That is, the penetration
of a Skyrmion may be viewed as the annihilation of the incoming
Skyrmion on the wall and the subsequent creation of a Skyrmion on
the other side. However, the annihilation results in traveling waves
along the wall that carry off a bit of the coherence required to
produce a Skyrmion on the other side. Hence, even though there is
enough energy in the vicinity of the collision, a Skyrmion is
unable to be created on the other side of the wall. We think that 
these considerations apply equally well to monopole-wall interactions
and that monopoles will - for all cosmological purposes - never 
penetrate the wall

%\smallskip

(iii) Energetically, the most favored state is where the monopole
unwinds. Even if this does not happen in a single monopole-wall
collision, it will occur if there are several interactions. Given
that the walls are very efficient at sweeping the universe, 
multiple monopole-wall interactions can easily occur.

%\smallskip

(iv) The interactions of vortices and domain walls separating
the A and B phases of $^3$He have been studied and also observed 
experimentally. It is found that singular vortices do not penetrate 
from the B phase into the A phase\cite{trebin}.
%\end{enumerate}

Based on the arguments above, we conjecture that the monopoles are 
trapped on the walls and since they can unwind on the wall,
will spread out as traveling waves. Eventually, as we will see 
below, the percolated wall will collapse and all the monopoles will be
eliminated. (This will be simplest to see in a 
closed universe in which the total magnetic charge must vanish.)

% Dec. 15, 1997: new part...

In view of the fact that we have not proved that monopoles will unwind,
it is useful to derive constraints on the probability of unwinding
that will enable a successful resolution of the monopole problem.
Let $\xi (t)$ be the average wall separation at time $t$.  Then the
time for the wall to move a distance $\xi (t)$ is $\tau (t)=\xi (t) /v_w$.
We will further assume that the wall network at time $t_0+\tau (t_0)$
is completely independent of the network at time $t_0$ and provides
a ``fresh'' (uncorrelated) set of walls that sweep the universe. The 
number of ``correlation times'' between wall formation at time $t_f$ and 
wall decay at time $t_d$ is:
$
N= {\rm log}_2 [ {{(t_d - t_f)} / {\tau (t_f)}} ] \ .
$
Let $1-p$ denote the probability that a monopole will survive for
a correlation time. (This includes both the possibility that the 
monopole may not encounter a wall and that it may not unwind
after being hit by a wall.) We will assume that $p$ and 
$\alpha = t/\xi (t)$ are constants which is reasonable if the
wall network scales. Then the probability that a monopole will
survive until the time $t_d$ is $S = (1-p)^N$
and this is constrained by cosmology to be less than a critical
value $S_* \sim 10^{-12}$. Hence we need $p > 1-S_*^{1/N}$.
To derive a numerical estimate of the upper bound on $N$, 
we take $t_d \sim (G \sigma )^{-1}$ 
(the time at which walls would start dominating \cite{avps}) and 
$\tau (t_f) \sim M_P/v^2$, where $M_P \sim 10^{19}$ GeV is the Plank 
scale. This gives,
$
N \sim {\rm log}_2 [ M_P / \sqrt{\lambda '} v ] \ .
$
For $v /M_P = 10^{-4}$ and $\lambda ' =0.1$, $N$ is about 15.
Therefore, with these parameters, we need $p > 0.8$ for a cosmologically
acceptable number of monopoles to have survived.

After the monopoles have been swept up by the walls, we are left with 
walls that are continually straightening out \cite{wallevoln}. 
At a certain time, the
condition in eq. (\ref{tension}) is violated, and the walls start
collapsing due to the pressure difference between the two sides.
The walls will decay before they come to dominate the universe
provided \cite{avps}: $\Delta F_V  > G \sigma ^2 V$.
This gives an additional constraint on $\gamma$
$$
\gamma > {{10 G\sigma \sqrt{\lambda '}} \over {m}} 
        \sim {{10}\over {\sqrt{\lambda '}}} 
                \left ( {{M_G}\over {M_P}}\right ) ^2
$$
where, $M_G \sim 10^{14}$ GeV is the GUT scale. Once the walls
are pressure driven, the volume fraction in the $\Phi_+$ vacuum
will increase at the expense of the $\Phi_-$ vacuum. Finally 
all of space will be in the $\Phi_+$ vacuum and no domain walls
will remain.

Summarizing the strongest constraints on $\gamma$, we need,
$$
10^{-9} {\lambda '} ^{-1/2} \lesssim 
\gamma \lesssim 10 {\lambda '} ^{+1/2}
$$
for the domain walls to sweep away the monopoles and subsequently
decay safely. The lower bound comes from requiring that the
walls never dominate the universe and the upper bound comes
from requiring that the walls percolate and that there is a period
during which the wall evolution is tension dominated\footnote{
If the walls do not percolate, they will all be finite and
will collapse without sweeping through the whole 
volume of the universe. If the pressure term becomes important
before the monopoles have been swept away,
some of the monopoles that were formed in
the $\Phi_+$ vacuum, will not be swept up by the walls and
will survive. In these cases, depending on the fraction of space 
that remains unswept, we may or may not have a monopole problem.}.

The domain walls in the $\gamma =0$ case can also be eliminated
if the $Z_2$ symmetry is anomalous under a strongly-coupled
$SU(N)$ gauge group\cite{anomaly}. 
This can be the case if $\Phi$ gives mass to an odd
number of fermionic flavors transforming in the fundamental representation
of $SU(N)$. Then, if there is no extra matter charged under $SU(N)$,
the $Z_2$ symmetry is expected to be explicitly broken by instantons.
Provided the strong scale of $SU(N)$, $\Lambda$, is smaller than $M_G$,
and the Yukawa coupling constants of fermions are of order one,
the instanton-induced energy difference between the $\pm$ vacua
is $\sim \Lambda^4$.  This bias will only turn on at a time 
$t_{anom}= {M_P}/ {\Lambda^2}$
and the requirement that walls never dominate the universe 
leads to the bound, $ \Lambda^2 > {M_G^3}/ {M_P}$.
(This rules out QCD as the source for an explicit bias).

Yet another source of bias\cite{goranrai} can be a higher dimensional 
gravity-induced Planck scale suppressed operators that, on general grounds, 
are expected not to respect the global symmetries of the theory.

This resolution of the monopole problem frees inflation from
having to occur during or after the GUT phase transition. 
Our considerations also show that the interactions of various
defects produced during a phase transition can be vital to 
cosmology. In particular, one class of defects can erase another
class.

{\it Acknowledgements:} 
GD would like to thank Savas Dimopoulos for discussions. TV is grateful 
to Alan Guth, Rich Holman, Misha Shaposhnikov, Mark Trodden and Alex Vilenkin
for helpful comments, to Grisha Volovik for explaining the
interactions of defects in $^3$He and providing references,
and, to the DoE for support.

\end{document}